\documentstyle[pra,aps]{revtex}
\begin{document}
\draft
\title{Beyond the Thomas-Fermi approximation for a trapped condensed
 Bose-Einstein gas}
\author{Alexander L.~Fetter}
\address{Departments of Physics and Applied Physics, Stanford University,
Stanford, CA 94305-4060 U.S.A.}
\author{David L.~Feder\footnote{Present address: Electron and Optical Physics
Division, National Institute of Standards and Technology, Gaithersburg, MD
20899.}}
\address{Department of Physics and Astronomy, McMaster University, Hamilton,
Ontario L8S 4M1, Canada}
\date{\today}
\maketitle

\begin{abstract}

Corrections to the zero-temperature Thomas-Fermi description of a dilute
interacting condensed Bose-Einstein gas confined in an isotropic harmonic
trap arise due to the presence of a boundary layer near the condensate
surface. Within the Bogoliubov approximation, the various contributions to the
ground-state condensate energy all have terms of order $R^{-4}\ln R$ and
$R^{-4}$, where $R$ is the number-dependent dimensionless condensate radius in
units of the oscillator length $\sqrt{\hbar/m\omega_0}$. The zero-order
hydrodynamic density-fluctuation amplitudes are extended beyond the
Thomas-Fermi radius through the boundary layer to provide a uniform
description throughout all space. The first-order correction to the excitation
frequencies is shown to be of order $R^{-4}$.

\end{abstract}
\pacs{03.75.Fi, 05.30.Jp, 32.80.Pj}
\section{introduction}

The recent experimental realization of Bose-Einstein condensation (BEC) of
alkali-metal gases in magnetic traps~\cite{cornell1,ketterle1,hulet} has
generated great interest in the physics of a confined, interacting,  dilute
Bose gas.  In the Bogoliubov approximation~\cite{bogoliubov}, which is
generally valid at temperatures well below the BEC
transition temperature, the macroscopic occupation of the ground state far
exceeds that of excited states; the condensate is then described by the
relevant Gross-Pitaevskii (GP) equation~\cite{gross,pit}. This equation
has been solved numerically for bosons in both
isotropic~\cite{edwards1,ruprecht,edwards2}
and anisotropic~\cite{dalfovo1} traps, and analytically in the
limits of both
a nearly ideal gas (weak interparticle interactions) and a dilute
nonideal gas
(strong interparticle interactions)~\cite{baym,fetter1}.

Given the
resulting condensate wave function $\Psi$,   the linearized
small normal-mode
 amplitudes
$u$ and $v$ satisfy the coupled Bogoliubov equations \cite{pit},  which
contain the condensate
density $|\Psi|^2$ through the potential energy
 of interaction.
An alternative hydrodynamic approach makes use of the fluctuations in
the density and the velocity.  It has  been shown to be wholly equivalent
to the  Bogoliubov description~\cite{fetter2,wu,fetter3}, and it accurately
 describes  the  low-lying excited states~\cite{stringari,EDCB,Rup}, as
  recent  experiments have verified \cite{cornell2,ketterle2} in
considerable detail.

For large particle number, the mean kinetic energy of the condensate is
much smaller than both the interaction (Hartree) and trap confinement
energies. Neglecting the kinetic energy entirely, corresponding to the
Thomas-Fermi (TF)
approximation, provides an accurate description of the condensate in the
interior of the cloud. Near the surface of the trapped gas, however, the
kinetic and external potential energies become comparable, and the TF
approximation breaks down. Using a boundary-layer theory, Dalfovo
{\it et al.}~\cite{dalfovo2} have calculated the kinetic energy as a
function of
the condensate radius;  the formally divergent TF kinetic energy is
cut off by
a boundary layer of thickness
$\delta\propto R^{-4/3}$, where $R$ is the (large) dimensionless TF
condensate radius.

The present work extends the boundary-layer formalism of
Ref.~\cite{dalfovo2} to determine the leading-order corrections
to
the TF description of the condensate wave function, the condensate
energy, and the low-lying collective modes. Section II summarizes the
basic formalism and obtains the first correction  to  the condensate wave
function, both in the bulk (of order $R^{-4}$) and in the boundary layer
(of order
$\delta$).   In Sec.~III, we show that this
perturbative  expansion  gives rise to
correction  terms of relative order
$R^{-4}\ln R$ and $R^{-4}$ in
the normalization of the condensate wave function and in   the
trap and  interaction  energies of the condensate. A combination
with the previously evaluated kinetic energy \cite{dalfovo2} provides the
leading   correction to the TF total  condensate energy.  In  Sec.~IV, we
show that the structure of the condensate boundary layer  plays an essential
role in constructing the corresponding boundary layer for  the analytical
hydrodynamic normal modes
\cite{stringari};   the resulting hydrodynamic amplitudes for the density
and current fluctuations  vanish exponentially as $r\to \infty$, as
expected from their equivalence to the  eigenfunctions of the coupled
Bogoliubov equations.

\section{condensate wave function}

For a dilute interacting inhomogeneous Bose gas in an isotropic trap
potential $V_{\rm
ext}$ at zero temperature, the total occupation of the  excited
states is small, and one can
apply the Bogoliubov approximation~\cite{bogoliubov,fetter4}. The
spatially varying
condensate wave function $\Psi({\bf r})$ is then isotropic and satisfies
the (stationary)
Gross-Pitaevskii~\cite{gross} equation:
\begin{equation}
(T+V_{\rm ext}+V_{\rm H}-\mu)\Psi(r)=0,
\label{GP0}
\end{equation}

  \noindent where $T=-\hbar^2\nabla_r^2/2m$ is the kinetic energy, and the
trap potential $V_{\rm ext}(r)=m\omega_0^2r^2/2$ is taken to be isotropic
for simplicity. The Hartree energy is the mean energy of interaction of a
particle at $\bf r$ with all the other particles, defined as
$V_{\rm H}(r)=\int d^3 r'\,U({\bf r-r'}) \,|\Psi({\bf r}')|^2
\approx g|\Psi(r)|^2$, where the last form reflects the
 short-range two-body interaction $U({\bf r})\approx g\delta^{(3)}({\bf
r})$. To leading order, the coupling constant
$g=4\pi a\hbar^2/m$ is written in terms of the (low-energy)
$s$-wave scattering length $a$
 to make contact with experiment; in the present work we only consider
$a>0$, corresponding to interparticle repulsion. The chemical potential
$\mu$
fixes the total number of condensed atoms
$N_0=\int\,d^3r\,|\Psi({\bf r})|^2$
through
\begin{equation}
\mu N_0=\langle T\rangle+\langle V_{\rm ext}\rangle+\langle V_{\rm
H}\rangle,
\label{n0_def}
\end{equation}

\noindent where the noncondensate contribution to the chemical potential is
neglected and $\langle\cdots\rangle \equiv \int d^3r\,\Psi^*\cdots\Psi$
denotes an
expectation value in the condensate ground state.

It is convenient to use the the oscillator length
$d_0=\sqrt{\hbar/m\omega_0}$ and the oscillator energy
$\hbar\omega_0$ as the basic dimensional  units.  Thus we introduce the
dimensionless length
$z\equiv r/d_0$,
 and the
dimensionless parameter
\begin{equation}\eta_0=\frac{N_0a}{d_0}\end{equation}
 that characterizes the strength of
the interparticle interactions. Experimental conditions typically give
$\eta_0\gg 1$, corresponding to a strongly interacting system~\cite{baym}
(but the
Bogoliubov approximation still requires a dilute system with a large
interparticle spacing $n^{-1/3}$
relative to $a$). In  this TF limit, the repulsive interactions expand the
condensate
cloud beyond
$d_0$ to a large dimensionless radius $R\propto N_0^{1/5}$ (expressed in
units of
$d_0$)~\cite{baym}.  The kinetic energy in Eq.~(\ref{GP0}) then becomes
small, and the condensate density has a simple parabolic form $n_0=|\Psi|^2
\approx g^{-1}(\mu - V_{\rm ext}) = (\mu/g)(1-z^2/R^2)$.  Since
$\eta_0\propto N_0$ scales like
$R^5$, it is convenient to define
\begin{equation}\eta_0 \equiv \tilde\eta_0 R^5,\end{equation}
where $\tilde\eta_0$ approaches a constant value for large $R$.  Defining a
scaled variable
$x=z/R$, the full Gross-Pitaevskii equation~(\ref{GP0}) becomes:
\begin{equation}
\big[-\case1/2 \epsilon\nabla_x^2+\case1/2 x^2+|\tilde\Psi({x})|^2
-\tilde\mu\big]\tilde\Psi({ x})=0,
\label{GP}
\end{equation}

\noindent where $\epsilon=1/R^4$ is a small coefficient in the present
limit
 $R\rightarrow\infty$ and $\tilde{\mu}=\mu/R^2$ is the scaled chemical
potential.  Here, the  scaled condensate wave function
\begin{equation}|\tilde{\Psi}|^2\equiv\frac{4\pi
d_0^3\eta_0}{N_0R^2}\,|\Psi|^2=\frac{4\pi
d_0^3R^3\tilde\eta_0}{N_0}\,|\Psi|^2\label{Psisq}\end{equation}
 becomes independent of $R$   for large $R$.

The normalization of the condensate wave function is then
written
\begin{equation}
\tilde{\eta}_0\equiv{\eta_0\over R^5}=
\int_0^{\infty}dx\,x^2|\tilde{\Psi}(x)|^2,
\label{normal}
\end{equation}

\noindent which defines the condensate radius in terms of
the particle number. Correspondingly, the scaled chemical potential follows
>from  Eq.~(\ref{n0_def}) as $\tilde\mu =\langle \tilde V_{\rm ext}+\tilde
V_H +
\tilde T\rangle$, where
\begin{equation}
\langle\tilde{V}_{\rm ext}\rangle\equiv{\langle V_{\rm ext}\rangle
\over R^2}
={N_0\over
2\tilde{\eta}_0}\int_0^{\infty}dx\,x^4\left|\tilde{\Psi}(x)\right|^2;
\label{extern}
\end{equation}
\begin{equation}
\langle\tilde{V}_{\rm H}\rangle\equiv{\langle V_{\rm H}\rangle\over R^2}
={N_0\over\tilde{\eta}_0}\int_0^{\infty}dx
\,x^2\left|\tilde{\Psi}(x)\right|^4;
\end{equation}
\begin{equation}
\langle\tilde{T}\rangle\equiv{\langle T\rangle\over R^2}={N_0\epsilon\over
2\tilde{\eta}_0}\int_0^{\infty}dx\,x^2\left|\tilde{\Psi}'(x)\right|^2,
\label{kinetic}
\end{equation}

\noindent where all the energies are expressed in units of $\hbar\omega_0$, and
$\tilde{\Psi}'=d\tilde{\Psi}/dx$ is the scaled radial derivative.

In the Thomas-Fermi (TF) limit, one sets the small parameter
 $\epsilon$ to zero
in Eq.\ (\ref{GP}). The approximate condensate wave function
$\tilde{\Psi}(x)\approx\tilde{\Psi}_{\rm TF}(x)\theta(1-x)=\sqrt{\frac{1}{2}(1
-x^2)}\theta(1-x)$ accurately describes the condensate in the
bulk~\cite{dalfovo1,baym}, where we take $\tilde\mu = \frac{1}{2}$,
{\it defining\/}
 the condensate radius by $R(\mu)= \sqrt{2\mu}$. With this choice, the
normalization condition Eq.~(\ref{normal}) then determines the condensate
number $N_0$ in terms of  the radius
$R$ and the chemical potential.

The Thomas-Fermi approximation fails near the surface at $x=1$, however, where
$\tilde{V}_{\rm ext}=\frac{1}{2}x^2$ is comparable to $\tilde{\mu}$; in this
region, the kinetic term in Eq.\ (\ref{GP}) becomes significant, giving rise
to a logarithmic divergence in Eq.~(\ref{kinetic}). For small positive
$\epsilon$, a boundary layer of thickness $\delta$ forms in the
vicinity of
$x=1$  where the condensate wave function varies rapidly. Using standard
techniques
in boundary-layer theory~\cite{bender}, we define an outer solution
$\tilde{\Psi}_{\rm outer}
\equiv\chi(x)$ valid  in the bulk region $0\leq x\leq x_0< 1$, and an inner
 solution
$\tilde{\Psi}_{\rm inner}$, valid throughout the surface region
$x_0\leq x\leq\infty$; an asymptotic analysis matches these two solutions
near the boundary
$x\approx x_0$.

The outer (bulk) solution $\chi$ may be expressed as a perturbation series
in powers of
$\epsilon$:
\begin{equation}
\chi(x)=\chi_0(x)+\epsilon\chi_1(x)+\cdots, \quad 0\leq x\leq x_0.
\label{outer}
\end{equation}

\noindent Substituting Eq.~(\ref{outer}) into Eq.~(\ref{GP}) yields

\begin{eqnarray}
O(\epsilon^0):&\quad&\chi_0^2(x)={\case1/2}(1-x^2);\label{chi0} \\
O(\epsilon^1):&\quad&2\chi_0(x)\chi_1(x)=
\frac{\nabla_x^2\chi_0}{2\chi_0 }={x^2-{\case3/2}\over(x^2-1)^2},\label{chi1}
\end{eqnarray}

\noindent where we have set $\tilde\mu = \frac{1}{2}$ and specialized to
the present case of a
real isotropic condensate wave function.  In order to
determine the asymptotic behavior of the outer solution near the boundary,
 one
may write $x=1+\delta X$ with $\delta|X|\ll 1$ and $|X|\gg 1$ (where $X$ is
large
and negative). As $X\rightarrow -\infty$, a straightforward calculation
 reveals
\begin{equation}
\chi(X)\sim\delta^{1/2}(-X)^{1/2}\left(1+{1\over 8X^3}\right)
-{\delta^{3/2}(-X)^{3/2}\over 4}\left(1-{21\over 8X^3}\right)+\cdots,
\label{asymptotic}
\end{equation}

\noindent where the leading-order behavior ($\propto \delta^{1/2}$) agrees
with previous calculations~\cite{lundh}.

The asymptotic behavior of the outer solution implies that the inner
solution has the form
$\tilde{\Psi}_{\rm inner}(X)\equiv\delta^{1/2}\Phi(X)$;  it  may be
expanded as a
series in $\delta$:
\begin{equation}
\tilde{\Psi}_{\rm inner}(X)=\delta^{1/2}\left[\Phi_0(X)+\delta\Phi_1(X)
+\cdots
\right],
\label{inner}
\end{equation}

\noindent where $X_0\leq X<\infty$
 and $X_0$ is large and negative. Substituting
Eq.~(\ref{inner}) into Eq.~(\ref{GP})  immediately gives the
``distinguished limit''~\cite{bender} $\epsilon=2\delta^3$ that balances
the leading-order (gradient and nonlinear) terms in the resultant
differential equation. We readily obtain:

\begin{eqnarray}
O(\delta^0):&\quad&\Phi_0''(X)-\left[X+\Phi_0^2(X)\right]\Phi_0(X)=0;
\label{inner0} \\
O(\delta^1):&\quad&\Phi_1''(X)-\left[X+3\Phi_0^2(X)\right]\Phi_1(X)=
-2\Phi_0'(X)+\case1/2 X^2\Phi_0.
\label{inner1}
\end{eqnarray}

\noindent It is straightforward to verify that, for $X\rightarrow -\infty$, the
inner functions $\Phi(X)$ exactly reproduce Eq.\ (\ref{asymptotic}):

\begin{eqnarray}
\Phi_0(X)&\sim&(-X)^{1/2}\left(1+{1\over 8X^3}\right);\label{a_inner0} \\
\Phi_1(X)&\sim&-{(-X)^{3/2}\over 4}\left(1-{21\over 8X^3}\right).
\label{a_inner1}
\end{eqnarray}

\noindent For large positive $X$, $\tilde{\Psi}_{\rm inner}(X)$ is
asymptotically proportional to an Airy function and therefore decays
exponentially. Indeed, since Eq.\ (\ref{inner0}) defines what is known
as the second Painlev\'e transcendent, the boundary condition
$\Phi_0(X\rightarrow\infty)\sim\sqrt{2}\hbox{Ai}(X)\sim
(2\pi)^{-1/2}\,X^{-1/4}\,\exp(-\frac{2}{3}X^{3/2})$ must be imposed
in order to
ensure an unbounded solution for $\Phi_0(X)$ with no critical points
over all
$X$~\cite{painleve}.

\section{condensate energies}

The normalization of the condensate wave function can now be
determined explicitly from Eq.~(\ref{normal}) by separating the integral
into two parts at $x_0$. An expansion of each contribution shows that the
resulting sum  is independent of the   matching point $x_0 =
1+\delta X_0$ and yields
\begin{equation}
\tilde{\eta}_0\approx{1\over 15}+\delta^2I+{1\over 8}\epsilon\,\ln\epsilon
+\epsilon\left({1\over 24}-{\ln 2\over 2}+J\right),\label{norm1}
\end{equation}

\noindent where $I$ and $J$ are definite integrals involving the inner
function
$\Phi_0(X)$ (see Appendix~\ref{appendix1} for details). In fact, $I$ can be
shown to
vanish identically, so that the leading corrections are of order
$R^{-4}\ln R$ and $R^{-4}$ (instead of order $R^{-8/3}$ implicit in  the
term
of order $\delta^2$).

A similar analysis yields the physical
quantities in Eqs.~(\ref{extern})-(\ref{kinetic}):
\begin{equation}
\langle\tilde{V}_{\rm ext}\rangle\approx{N_0\over 2\tilde{\eta}_0}\left[
{1\over 35}+\delta^2I+{1\over 24}\epsilon\,\ln \epsilon+\epsilon\left(
{5\over 24}-{\ln 2\over 6}+K\right)\right],
\label{extern2}
\end{equation}
\begin{equation}
\langle\tilde{V}_{\rm H}\rangle\approx{N_0\over\tilde{\eta}_0}\left[{2\over
105}
+{1\over 12}\epsilon\,\ln \epsilon+\epsilon\left(-{\ln 2\over 3}
+{L\over 2}\right)\right],
\label{hartree2}
\end{equation}
\begin{equation}
\langle\tilde{T}\rangle\approx -{N_0\over\tilde{\eta}_0}\left[{1\over 24}
\epsilon\,\ln \epsilon+\epsilon\left({1\over 12}-{\ln 2\over 6}
+{M\over 2}\right)\right],
\label{kinetic2}
\end{equation}

\noindent where the constants  $K$, $L$, and $M$ are also definite
integrals involving the inner function $\Phi_0(X)$.  A combination of
analytical and numerical techniques gives the explicit results

\begin{eqnarray}
I&=&0;\\
J&=&{\case3/4}L-{\case5/{24}};\\
K&=&{\case1/4}L-{\case7/{24}};\\
L&\approx&0.4539;\\
M&=&{\case1/2}L-{\case1/{12}},
\end{eqnarray}

\noindent such that $K+L=J+M$.

Inserting the explicit
expressions from Eqs.~(\ref{extern2})-(\ref{kinetic2}) into
Eq.~(\ref{n0_def}), we
immediately  recover
$\mu=\frac{1}{2}R^2$, demonstrating the internal consistency of the
calculation.  Equation (\ref{norm1}) implies
\begin{equation}
\eta_0(R)=\left({a\over d_0}\right)N_0\approx{R^5\over 15}-{R\over
2}\bigg[\ln \bigg(\frac{R}{A}\bigg) +\frac{7}{12}\bigg]\approx
{R^5\over 15}-{R\over
2}\ln
(1.4128\,R),\label{number}
\end{equation}

\noindent where the constant $A$ is
\begin{equation}
A=\case1/2 \exp\left({\case3/2 L+\case1/4}\right)\approx 1.268.
\end{equation}

\noindent This equation (\ref{number})  relates the condensate number to
the chemical potential (and hence the condensate radius);   conversely, its
inverse
\begin{equation}
R(\eta_0)\approx\left(15\eta_0\right)^{1/5}+\case 3/{10}(15\eta_0)^{-3/5}
\ln\left(84.46\,\eta_0\right)\label{R}
\end{equation}

\noindent
 relates the radius of the cloud (and the chemical potential) to the
condensate number.
 In each case, the
first terms correspond to the TF result~\cite{baym}.

The various contributions to the total
energy [Eqs.~(\ref{extern})-(\ref{kinetic})] yield
\begin{equation}
{\langle V_{\rm ext}\rangle\over N_0}\approx{3R^2\over 14}
\left[1+{5\over 3R^4}
\ln\left({R\over A}\right)\right];
\label{extern3}
\end{equation}
\begin{equation}
{\langle V_{\rm H}\rangle\over N_0}\approx{2R^2\over 7}\left[1-
{10\over R^4}\ln
\left({R\over A}\right)\right];\label{Hartree3}
\end{equation}
\begin{equation}
{\langle T\rangle\over N_0}\approx{5\over 2R^2}\ln\left({R\over A}\right),
\label{kinetic3}
\end{equation}

\noindent and the energies are expressed in units of $\hbar\omega_0$.

The
expression for the kinetic energy in Eq.~(\ref{kinetic3}) is a small
correction that involves only the leading contributions to the bulk
condensate wave function and the  boundary layer;  it reproduces the
result of Dalfovo {\it et al.}~\cite{dalfovo2}, and is similar to that
found in~\cite{lundh}. It should be kept in mind, however, that the present
calculation defines the condensate radius $R$  in terms of the
chemical potential $\mu\equiv{\case1/2}R^2$; it therefore includes
correction terms beyond the TF approximation, as shown in Eq.~(\ref{R}).

The remaining contributions Eqs.~(\ref{extern3}) and (\ref{Hartree3})
explicitly require the leading corrections $\chi_1$ and $\Phi_1$ and have not
been evaluated previously. The dimensionless total energy per particle
$\langle E\rangle/N_0 =\langle T+V_{\rm ext}+{\case1/2}V_{\rm H}\rangle$ is
found to be
\begin{equation}
{\langle E\rangle\over N_0}\approx{5R^2\over 14}\left[1+{4\over R^4}\ln
\left({R\over A}\right)\right],\label{Etot}
\end{equation}

\noindent where the first and second terms are, respectively, the standard TF
result and the combined contribution from both  the boundary layer and the
first correction to the bulk condensate wave function.  Equations
(\ref{number}) and (\ref{Etot}) together constitute a parametric
representation of the relation $\langle E(N_0)\rangle$ \cite {gonzalez}, and
it is not difficult to verify that $\mu = d\langle E\rangle/dN_0$.

In fact, Stringari~\cite{stringari,stringari2} has noted that
Eqs.~(\ref{extern3})-(\ref{Hartree3}) and (\ref{Etot}) can be obtained
solely from the expression for the kinetic energy (\ref{kinetic3}).  The
virial theorem implies that the
three contributions to the total energy must satisfy the condition
\begin{equation}
2\langle T\rangle+{\case3/2}\langle V_{\rm H}\rangle
-2\langle V_{\rm ext}\rangle=0.\label{virial}
\end{equation}

\noindent The leading-order contributions to the potential
energies $\langle V_{\rm ext}\rangle $ and $\langle V_{\rm H}\rangle $
are known from TF theory, and  the correction terms must have the same
$R$-dependence as the kinetic energy but with unknown coefficients.
Equation~(\ref{virial}) and the condition
$\mu=\partial\langle E\rangle/\partial N_0$ together determine for the
unknown coefficients, reproducing the expressions found above by direct
integration.

As shown in Fig.~\ref{figure1}, the results of the boundary-layer theory
agree  strikingly  with exact results obtained by
integrating the GP equation~(\ref{GP}) numerically. The analytical slow
$R^{-2}\ln R$ decay of the average kinetic energy per particle (shown dashed
bold) with the universal scaling parameter
$\eta_0=N_0a/d_0$ accurately captures the behavior obtained
numerically (shown solid bold). As a result, the boundary-layer theory
provides a much better estimate of the total energy per particle (and
therefore the chemical potential) than does the TF approximation. In spite
of the slow decrease in the magnitude of
$\langle T\rangle/N_0$ with the number of particles, however, it should be
emphasized that the TF approximation to the total energy and the chemical
potential is correct to better than
$1\%$ when
$\eta_0\sim 1000$, due to their $R^2$ increase.

\section{excited-state wave functions}

In the Bogoliubov approximation at zero temperature, the interparticle
repulsions excite only a small fraction  of all the
particles  out of the ground state into self-consistent normal modes. The
resulting eigenfunctions $u_j({\bf x})$ and $v_j({\bf x})$  and
eigenvalues
$E_j$ for the
noncondensate modes satisfy the coupled linear Bogoliubov
equations~\cite{bogoliubov,pit,fetter4}. For the present purpose, it is
convenient to factor out the exact
(real) GP condensate wave function $\Psi$, defining  sum and difference
amplitudes\cite{fetter3}
\begin{equation}
 F_j({\bf x})\equiv \frac{u_j({\bf x})+v_j({\bf x})}
{\Psi({\bf x})}\qquad\hbox{and}\qquad G_j({\bf x}) \equiv
\frac{u_j({\bf x})-v_j({\bf x})}{\Psi({\bf x})},\label{FG}
\end{equation}

\noindent which are essentially  the hydrodynamic amplitudes. Specifically,
the perturbation in the velocity potential $\phi_j$ is proportional to
$F_j$ (so that the perturbation in the current density is
proportional to
$\Psi^2 \bbox{\nabla}F_j$) and the density perturbation $\rho_j$ is
proportional to $\Psi^2G_j$.  In the TF limit, it is convenient to
rescale these amplitudes, with
$F_j = R\,\tilde F_j$ and
$G_j = \tilde G_j\,/R$,  yielding the coupled linear equations

\begin{eqnarray}
-\case1/2\bbox{\nabla\cdot}\big(\tilde\Psi^2\bbox{\nabla }\tilde
F_j\big)& = E_j\tilde \Psi^2\tilde G_j,\label{F}\\ \noalign{\smallskip}
2\tilde\Psi^4\tilde G_j
-\case1/2\epsilon\bbox{\nabla\cdot}\big(\tilde\Psi^2\bbox{\nabla}
\tilde G_j\big)&=E_j\tilde\Psi^2\tilde F_j,\label{G}
\end{eqnarray}

\noindent where $\tilde\Psi^2$ is the (real) scaled condensate density
from Eq.~(\ref{Psisq}). These Eqs.~(\ref{F}) and (\ref{G})
are self-adjoint  with a  normalization integral $ \int
d^3x\,\tilde\Psi^2\big(\tilde F_j^*\tilde G_j +
\tilde G_j^*\tilde F_j\big) = 1$ chosen to ensure positive energies $E_j$
for the stable solutions \cite{fetter4}.

In the TF regime ($R\to\infty$), Stringari \cite{stringari} has solved
these coupled equations   for a stationary isotropic
condensate, but the resulting hydrodynamic eigenfunctions are defined
only  inside the condensate ($x\le 1$).  In contrast, the original
Bogoliubov equations are well-defined throughout the whole region,
including the classically forbidden region ($x\gg 1$).  Thus it is
interesting to develop a boundary-layer description of the excited
states similar to that for the condensate.

\subsection{Outer (bulk) region}

In the ``outer'' region
($0\le x\le x_0<1$), we follow the procedure for the condensate and expand
the outer scaled amplitudes~(\ref{FG}) and eigenvalues
 in powers of $\epsilon$:
\begin{eqnarray}
\tilde{F}_j&\approx&\tilde{F}_j^0+\epsilon\tilde{F}_j^{\epsilon}+\ldots;\\
\tilde{G}_j&\approx&\tilde{G}_j^0+\epsilon\tilde{G}_j^{\epsilon}+\ldots;\\
{E}_j&\approx&{E}_j^0+\epsilon{E}_j^{\epsilon}+\ldots.\label{E_outer}
\end{eqnarray}

\noindent Substituting these expressions into Eqs.~(\ref{F}) and (\ref{G}),
and including the outer expansion of the condensate given in
Eq.~(\ref{outer}), one obtains:
\begin{eqnarray}
O(\epsilon^0):&\quad&
-\bbox{\nabla\cdot}\left(\chi_0^2\bbox{\nabla}\tilde{F}_j^0\right)
-\left(E_j^0\right)^2\tilde{F}_j^0=0\qquad;\qquad\tilde{G}_j^0
={E_j^0\tilde{F}_j^0\over 2\chi_0^2};\label{FG0_eq}\\
\noalign{\vskip0.1in}
O(\epsilon^1):&\quad&
-\bbox{\nabla\cdot}\left(\chi_0^2\bbox{\nabla}\tilde{F}_j^{\epsilon}
\right)-\left(E_j^0\right)^2\tilde{F}_j^{\epsilon}
={\left(E_j^0\right)^2\over
4\chi_0^2}\bbox{\nabla\cdot}\left[\chi_0^2\bbox{\nabla}
\left({\tilde{F}_j^0\over\chi_0^2}\right)\right]+2\bbox{\nabla\cdot}
\left(\chi_0\chi_1\bbox{\nabla}\tilde{F}_j^0\right)+2E_j^0E_j^{\epsilon}
\tilde{F}_j^0;
\label{F1_eq} \\
&\quad&\tilde{G}_j^{\epsilon}={E_j^0\over
8\chi_0^4}\bbox{\nabla\cdot}\left[\chi_0^2
\bbox{\nabla}\left({\tilde{F}_j^0\over\chi_0^2}\right)\right]+{1\over
2\chi_0^2}
\left(E_j^0\tilde{F}_j^{\epsilon}+E_j^{\epsilon}\tilde{F}_j^0\right)
-{E_j^0\chi_1\tilde{F}_j^0\over\chi_0^3}.\label{G1_eq}
\end{eqnarray}

 The spherical symmetry of the condensate density permits a
decomposition into spherical harmonics, with the normal-mode amplitudes
proportional to $Y_{lm}(\theta,\phi)$. Equations (\ref{FG0_eq}) may then be
solved explicitly, and the corresponding (unnormalized) zeroth-order radial
amplitudes can be taken as
\begin{equation}
\tilde F_{nl}^0(x)\equiv\rho_{nl}^0(x)=x^lP_{nl}(x^2)\quad;\quad
\tilde{G}_{nl}^0={E_{nl}^0\tilde{F}_{nl}^0\over 2\chi_0^2},\label{FG0}
\end{equation}

\noindent where the radial quantum number $n$ denotes the number of
nodes. Here the eigenvalues take the well-known result\cite{stringari}
\begin{equation}
E_{nl}^0=\sqrt{l+n\,(2n+2l+3)},
\end{equation}

\noindent and $P_{nl}(x^2) = F(-n, n+l+\case3/2; l+\case3/2; x^2)$ is a
hypergeometric function\cite{abramowitz} that terminates to give a polynomial
of order $x^{2n}$.

While analytical solutions to the inhomogeneous differential
equations~(\ref{F1_eq}) and (\ref{G1_eq}) are not easy to find, the asymptotic
behavior of all the outer solutions in the vicinity of the boundary layer
$x\sim 1$ may be readily ascertained. Setting $x=1+\delta X$ with $X\ll
-1$ and expanding through  $O(\delta^3)$, we obtain:
\begin{eqnarray}
O(\epsilon^0):&&\quad\tilde{F}^0_{nl}\sim\rho_{nl}^0(1)
+\delta X{\rho_{nl}^0}'(1)+{\case1/2}\delta^2X^2{\rho_{nl}^0}''(1)
+{\case1/6}\delta^3X^3{\rho_{nl}^0}'''(1); \label{F0_asympt} \\
&&\quad\tilde{G}^0_{nl}\sim-{E_{nl}^0\over 2\delta X}\Big\{G_a^0+\delta XG_b^0
+\delta^2X^2G_c^0+\delta^3X^3G_d^0\Big\};\label{G0_asympt} \\
\noalign{\vskip0.1in}
O(\epsilon^1):&&\quad\tilde{F}^{\epsilon}_{nl}\sim
{F_a^{\epsilon}\over\delta^3X^3}+{F_b^{\epsilon}\over\delta^2X^2}
+{F_c^{\epsilon}\over\delta X}+F_d^{\epsilon}+F_e^{\epsilon}\ln(-\delta X)
+F_f^{\epsilon}\ln^2(-\delta X);\label{F1_asympt} \\
&&\quad\tilde{G}^{\epsilon}_{nl}\sim -{E_{nl}^0\over 2\delta X}\left\{
{G_a^{\epsilon}\over\delta^3X^3}+{G_b^{\epsilon}\over\delta^2X^2}
+{G_c^{\epsilon}\over\delta X}+G_d^{\epsilon}+G_e^{\epsilon}\ln(-\delta X)
+G_f^{\epsilon}\ln^2(-\delta X)\right\}-{E_{nl}^{\epsilon}\rho_{nl}^0(1)
\over 2\delta X},
\label{G1_asympt}
\end{eqnarray}

\noindent where the constants $F_{a-f}^{\epsilon}$ and
$G_{a-f}^{0,\epsilon}$ are relegated to Appendix~\ref{appendix2} for
clarity. The expansion (\ref{F0_asympt}) is the conventional Taylor series
of the zeroth-order amplitudes $\tilde{F}_{nl}^0(x)$ about $x=1$, where a
prime denotes a derivative evaluated at $x=1$. A straightforward
calculation following from the properties of the hypergeometric
function\cite{fetter3,abramowitz} shows that
\begin{equation}
\rho_{nl}^0(1)= (-1)^n\frac{\Gamma(l+ \case3/2)\,n!}{\Gamma(n+ l +
\case3/2)}\quad;\quad{\rho_{nl}^0}'(1)=\big(E_{nl}^0\big)^2\rho_{nl}^0(1)
\quad\ldots,
\end{equation}

\noindent with higher-order values listed in Appendix~\ref{appendix2}. Since
$\epsilon=2\delta^3$, the outer solutions in the region $x\sim 1$ are
respectively
$\tilde{F}_{nl}\approx\tilde{F}_{nl}^0+2\delta^3\tilde{F}_{nl}^{\epsilon}$ and
$\tilde{G}_{nl}\approx\tilde{G}_{nl}^0+2\delta^3\tilde{G}_{nl}^{\epsilon}$.

\subsection{Inner (boundary-layer) region}

The need for a boundary-layer  description of the eigenfunctions is
clear from the form of condensate density
$\tilde\Psi(x)^2 = \delta\Phi(X)^2$ and the gradient
$\bbox{\nabla}_x = \delta^{-1}\bbox{\nabla}_X$ (arising from the
substitution
$x = 1+\delta X$).  As a result,  the second term on the left-hand
side of Eq.~(\ref{G}) is now of order $\epsilon\times\delta/\delta^2\sim
\delta^2$, which is wholly comparable with the first term and thus no
longer negligible in zero order.  It is not difficult to see that the
proper scaling of the two inner amplitudes in the boundary layer is
\begin{equation}
\tilde F_{nl}(x) = A_{nl}(X)\qquad\hbox{and}\qquad
\tilde G_{nl}(x) = \frac{B_{nl}(X)}{\delta}.\end{equation}

\noindent The gradients in Eqs.~(\ref{F}) and (\ref{G}) must be
expanded in ascending powers of
$\delta$, leading to the following coupled equations for the inner
solutions
\begin{equation}
-\frac{d}{dX}\bigg(\Phi^2{dA_{nl}\over dX}\bigg)
-\delta{2\Phi^2\over 1+\delta X}{dA_{nl}\over dX}
+\delta^2{l(l+1)\Phi^2\over(1+\delta X)^2}A_{nl}= 2\delta
E_{nl}\Phi^2B_{nl},\label{A}
\end{equation}
\begin{equation}
-\frac{d}{dX}\bigg(\Phi^2{dB_{nl}\over dX}\bigg)
-\delta{2\Phi^2\over 1+\delta X}{dB_{nl}\over dX}
+\delta^2{l(l+1)\Phi^2\over(1+\delta X)^2}B_{nl}+2\Phi^4B_{nl}
=E_{nl}\Phi^2A_{nl}.\label{B}
\end{equation}

The asymptotic expressions for the outer amplitudes in the boundary region,
Eqs.~(\ref{F0_asympt})-(\ref{G1_asympt}), indicate that a simple expansion of
the inner functions in powers of $\delta$ is insufficient. In order to ensure
a match in the vicinity of the boundary layer to order
$O(\epsilon)=O(\delta^3)$, the inner solutions must also include the
non-trivial terms $\delta^3\ln\delta$ and $\delta^3\ln^2\delta$.
The eigenfunctions and eigenvalues are therefore expanded in the form:
\begin{eqnarray}
A_{nl}(X) &=& A_{nl}^0(X)+\delta A_{nl}^1(X)+\delta^2A_{nl}^2(X)
+\delta^3A_{nl}^3(X)+\delta^3\ln\delta A_{nl}^4(X)
+\delta^3\ln^2\delta A_{nl}^5(X),\label{A_expand}\\
B_{nl}(X) &=& B_{nl}^0(X)+\delta B_{nl}^1(X)+\delta^2B_{nl}^2(X)
+\delta^3B_{nl}^3(X)+\delta^3\ln\delta B_{nl}^4(X)
+\delta^3\ln^2\delta B_{nl}^5(X),\label{B_expand}\\
E_{nl} &=& E_{nl}^0+\delta E_{nl}^1+\delta^2E_{nl}^2
+\delta^3E_{nl}^3+\delta^3\ln\delta E_{nl}^4+\delta^3\ln^2\delta E_{nl}^5.
\label{E_inner}
\end{eqnarray}

\noindent Inserting these expansions into Eqs.~(\ref{A}) and (\ref{B}), and
taking into account the inner expansion for the condensate
$\Phi(X) = \Phi_0(X) + \delta\Phi_1(X) +\cdots$, one obtains equations
for the lowest-order inner functions:
\begin{eqnarray}
O(\delta^0):&&\quad -\left(\Phi_0^2{A_{nl}^0}'\right)'=0;\label{A0:eq} \\
&&\quad -\left(\Phi_0^2{B_{nl}^0}'\right)'+2\Phi_0^4B_{nl}^0=E_{nl}^0\Phi_0^2
A_{nl}^0;\label{B0:eq} \\
\noalign{\vskip0.1in}
O(\delta^1):&&\quad -\left(\Phi_0^2{A_{nl}^1}'\right)'-2\left(\Phi_0
\Phi_1{A_{nl}^0}'\right)'-2\Phi_0^2{A_{nl}^0}'=2E_{nl}^0\Phi_0^2B_{nl}^0;
\label{A1:eq} \\
&&\quad -\left(\Phi_0^2{B_{nl}^1}'\right)'-2\left(\Phi_0\Phi_1{B_{nl}^0}'
\right)'-2\Phi_0^2{B_{nl}^0}'+2\Phi_0^4B_{nl}^1+8\Phi_0^3\Phi_1B_{nl}^0
\nonumber \\
&&\qquad\qquad =E_{nl}^0\left(\Phi_0^2A_{nl}^1+2\Phi_0\Phi_1A_{nl}^0\right)
+E_{nl}^1\Phi_0^2A_{nl}^0,\label{B1:eq}
\end{eqnarray}

\noindent where a prime denotes a derivative with respect to $X$.

Equation~(\ref{A0:eq}) can be solved by taking $A_{nl}^0$ as a constant, and
comparison with the first term of Eq.~(\ref{F0_asympt}) shows that
$A_{nl}^0=\rho_{nl}^0(1)$. In contrast, Eq.~(\ref{B0:eq}) is inhomogeneous,
with the explicit solution
\begin{equation}
B_{nl}^0(X)=-{E_{nl}^0\rho_{nl}^0(1)\over \Phi_0(X)}\,{d \Phi_0(X)\over dX}
=-\case1/2 E_{nl}^0\rho_{nl}^0(1)\,{d\ln \Phi_0(X)^2\over dX}.\label{B0_large}
\end{equation}

\noindent By inserting the asymptotic expression (\ref{a_inner0}) for
$\Phi_0(X)$, it is simple to verify that $B_{nl}^0(X)$ matches the leading term
of Eq.~(\ref{G0_asympt}).

The solution to Eq.~(\ref{A1:eq}) is readily found
to be $A_{nl}^1(X) = {\rho_{nl}^0}'(1)\,X$, which is identical to the second
correction term of the outer solution. The other
correction $B_{nl}^1(X)$ satisfies a somewhat more complicated inhomogeneous
equation (\ref{B1:eq}).  While a closed-form solution cannot be found, the
behavior in the overlap region $X\rightarrow -\infty$ is easily obtained:
\begin{equation}
B_{nl}^1(X)\sim -{E_{nl}^0\rho_{nl}(1)\over 2}\left({E_{nl}^0}^2
-{1\over 2}\right)-{E_{nl}^1\rho_{nl}(1)\over 2X}.
\end{equation}

\noindent Matching the inner solution $B_{nl}^1$ with the outer
function $\tilde{G}_{nl}^0$ to order $\delta$ requires $E_{nl}^1=0$. It then
becomes possible to write an expression for $B_{nl}^1(X)$ valid for all $X$:
\begin{equation}
B_{nl}^1(X)  = {E_{nl}^0\rho_{nl}^0(1)\Xi_a(X)
+ E_{nl}^0{\rho_{nl}^0}'(1)\Xi_b(X)\over \Phi_0(X)},
\end{equation}

\noindent where the two functions $\Xi_a$ and $\Xi_b$ satisfy inhomogeneous
equations similar to Eq.~(\ref{inner1}) for $\Phi_1$:
\begin{equation}
-\Xi_a''+(X+3\Phi_0^2)\Xi_a = 4\Phi_0\Phi_0'\Phi_1-2\Phi_0\,{d\over dX}
\bigg(X+{\Phi_1\over \Phi_0}\bigg)\,
{d\over dX}\bigg({\Phi_0'\over \Phi_0}\bigg),
\end{equation}
\begin{equation}
-\Xi_b''+ (X+3\Phi_0^2)\Xi_b =X\Phi_0.
\end{equation}
For large negative $X$, the solutions approach
$\Xi_a\sim {1\over 4}(-X)^{1/2}$ and $\Xi_b\sim -{1\over 2}(-X)^{1/2}$, so
that $B_{nl}^1$ indeed matches the first correction term in
Eq.~(\ref{G0_asympt}). Each additional term in $\delta$ may
be analyzed in a similar fashion; the procedure is straightforward but
extremely tedious, so explicit solutions for the inner functions in the
overlap region (which are lengthy) are omitted for brevity. It is important to
note, however, that {\em both} inner functions $A_{nl}$ and $B_{nl}$ be
properly matched to their outer region counterparts at each stage, in order to
yield conditions on both the unknown coefficients and the eigenvalue
corrections.

By following the above prescription in turn for each term in the inner
expansion, it can be shown that all corrections to the eigenvalue with
prefactors smaller than $O(\delta^3)$ (including the $O(\delta^3\ln\delta)$ and
$O(\delta^3\ln^2\delta)$ terms) must vanish identically. This result may be
formally understood as follows. The density perturbation and the
velocity-potential perturbation each have corrections to all orders both in
the inner and outer regions; in contrast, the only correction term in the
outer perturbation expansion of the energy~(\ref{E_outer}) is of order
$\delta^3$. In order to ensure a smooth asymptotic match between the bulk and
surface amplitudes to each successive order, all the energy corrections
appearing in the inner expansion (\ref{E_inner}) with coefficients smaller
than $\delta^3$ must match to zero (the order $\delta$ case was considered
explicitly above). The introduction of logarithmic terms to the outer
perturbation expansions would give rise to additional contributions in the
overlap region, leading to inconsistencies in the asymptotic match. The
eigenvalue correction of order $\delta^3$ is finite, however:
$E_{nl}^3=2E_{nl}^{\epsilon}$. Thus,
\begin{equation}
E_{nl}=E_{nl}^0+{E_{nl}^{\epsilon}\over R^4}.\label{E_shift}
\end{equation}

\noindent Eqs.~(\ref{R}) and (\ref{E_shift}) together yield the
number-dependence of the excitation frequencies in the TF limit. The
asymptotic match reveals an energy correction of order $\epsilon=R^{-4}$
exists, but unfortunately it does not indicate its magnitude nor its variation
with $n,l$.

The present boundary-layer theory indicates that the energy correction
of order $\delta^3\ln\delta$ vanishes identically.  In contrast, a sum-rule
approach~\cite{stringari,stringari3} does yield a logarithmic correction to the
eigenvalues, proportional to the ratio of the average kinetic energy
(\ref{kinetic3}) and external potential energy (\ref{extern3}). This latter
approach assumes that a single frequency exhausts the sum-rule. Such an
assumption is thought to be valid in the hydrodynamic regime where a given
perturbation excites essentially all of the atoms into a particular low-energy
collective mode~\cite{puff,stringari4}. In the vicinity of the boundary layer,
however, the density of atoms decreases considerably, and the single-mode
approximation may become insufficient. In practice, any logarithmic correction
to the energy eigenvalues would be experimentally or numerically detectable
only if the magnitude of $E_{nl}^{\epsilon}$ were strongly dependent upon $n$
and $l$, or if the number of atoms were very low (small $R$). One may estimate
the difference between $n=0$ energies obtained with and without a logarithmic
correction, by assuming 
$E_{0l}^{\epsilon}\approx l(l-1)\beta_l/2$~\cite{stringari} where
$\beta_l=\log R$ and $\beta_l=1$, respectively. The deviation between the two
approximations is independent of angular momentum, is largest when
$\eta_0\approx 30$, and is at most $2\%$ of the mode frequency when $l=10$. At
present, therefore, the data are consistent with either theory.

The $(n,l)$-dependence of the energy correction $E_{nl}^{\epsilon}$ in
Eq.~(\ref{E_shift}) can not be found by conventional perturbation theory.
This situation arises because the explicit integrals, which eliminate the
logarithmic divergences, contain the nominally unperturbed inner
functions;   these functions, in turn, are themselves solutions of
differential equations that implicitly include the perturbing terms.
Consequently, we have considered the readily derived variational expression
\begin{equation}
E_j = \frac{\int dV\,\big[{\textstyle{1\over
2}}\tilde\Psi^2\bbox{\nabla}\tilde
F_j^*\cdot\bbox{\nabla}\tilde F_j + 2\tilde\Psi^4\tilde
G_j^*\tilde G_j +{\textstyle{1\over
2}}\epsilon \tilde\Psi^2\bbox{\nabla}\tilde
G_j^*\cdot\bbox{\nabla}\tilde G_j\big]}{\int dV\,\tilde \Psi^2 \big(\tilde
G_j^*\tilde F_j + \tilde F_j^*\tilde G_j\big)},
\end{equation}
which is stationary for small variations about the exact solutions
$\tilde F_j$ and $\tilde G_j$.
As a trial function, we use the unperturbed outer solutions in
Eqs.~(\ref{FG0}), and the  corresponding two-term inner solutions given in
Eqs.~(\ref{A_expand}) and (\ref{B_expand}); taken together, these
expressions constitute  uniform unperturbed solutions throughout all
space.  If the various integrals are divided at a point $x_0 = 1 +\delta
X_0$,with $X_0\ll -1$ and $\delta|X_0|\ll 1$, the correction terms arising
from the behavior of the outer solution near the TF boundary precisely cancel
with those from the inner boundary-layer solutions, leaving dimensionless
integrals of order $\ln\delta$ through $\delta^3\ln^2\delta$. The asymptotic
match requires that only the term proportional to $\delta^3$ remains finite,
and a detailed evaluation shows that the integrals up to order
$\delta^2\ln\delta$ indeed vanish. The explicit calculation to order
$\delta^3$ is prohibitive, however, due to the profusion of relevant terms.

The boundary-layer solutions given above now suffice to determine the
approximate density fluctuation amplitude
$\rho_{nl} \propto \tilde\Psi^2\tilde G_{nl}$ to order $\delta$
throughout the whole physical region. Inside the condensate, away from the
boundary ($0\le x\le x_0<1$), the outer solution is simply the zero-order
polynomial $\rho_{nl}^0(x)$  found by Stringari\cite{stringari}. The uniformly
matching inner solution $\Phi(X)^2B_{nl}(X)$ in the interval
$X_0\leq X < \infty$ (where $x_0 = 1+\delta X_0$, with $\delta |X_0|\ll 1$ and
$X_0\ll -1$) reduces to
\begin{equation}
\rho_{nl}\approx
-\rho_{nl}^0(1)\,\frac{d\Phi_0(X)^2}{dX}
+2\delta{\rho_{nl}^0}'(1)\Phi_0(X)\Xi_b(X)
+2\delta\rho_{nl}^0(1)
\bigg[\Phi_0(X)\Xi_a(X)-2\Phi_1(X)\frac{d\Phi_0}{dX}\bigg].\label{rho}
\end{equation}
It is not difficult to verify that the third term in Eq.~(\ref{rho}) vanishes
for $X\to\pm\infty$, although it is nonzero inside  the boundary
region.  For large positive $X$, each term in Eq.~(\ref{rho}) vanishes
exponentially, so that the density fluctuations become negligible beyond
the surface region.

Figure~\ref{figure2} shows the spatial variation of the hydrodynamic
amplitudes in the inner (boundary-layer) region, including corrections of
order $\delta$ and $\delta^2$. The velocity potential perturbation
$\phi_{nl}(X)\propto A_{nl}(X)$ (dot-dashed line) to order $\delta$ or
$\delta^2$ is a linear or quadratic function of the inner coordinate $X$ and
therefore diverges at large positive $X$. In contrast, both the density
fluctuation $\rho_{nl}(X)\propto\Phi^2(X)B_{nl}(X)$ from Eq.~(\ref{rho})
(solid line) and the perturbation in the current density
$j_{nl}\propto\Phi^2(X)A_{nl}'(X)$ (short dashed line) vanish exponentially in
the limit $X\to\infty$. In fact, such behavior of the density and
current-density amplitudes for $X\to\infty$ holds to all orders in $\delta$
as a direct consequence of the condensate wave function's exponential
decay. The results of the boundary-layer theory differ from those obtained
within the TF approximation, where the density-fluctuation amplitudes are
merely finite at the TF radius.

The inner solutions provide a more detailed view of the behavior of these
amplitudes near the condensate surface. In particular, the velocity potential
and density perturbation coincide only in the outer region $X\ll -1$,
reflecting the fact that in the limit $\epsilon\to 0$ the zeroth-order
amplitudes $\tilde{F}_{nl}^0$ and $\chi_0^2\tilde{G}_{nl}^0$ obey the same
differential equation (\ref{FG0_eq}). Since these unperturbed outer amplitudes
are polynomials of order $2n+l$ in the variable $x=1+\delta X$, the inner
functions must be expanded to at least order $\delta^{2n+l}$ in order to ensure
a perfect asymptotic match.

\section{discussion}

The zero-temperature Thomas-Fermi description of an interacting dilute
Bose-Einstein gas confined in an isotropic harmonic trap has been extended to
include contributions from the condensate surface. For the  condensate wave
function, we have generalized the boundary-layer formalism of Dalfovo
{\it et al.\/}\cite{dalfovo2},
obtaining analytic expressions for the expectation values of the
trap and interaction energy  that  include  the leading corrections due
to the surface layer and to the bulk condensate wave function.  The
resulting total ground state energy, which  includes all terms of order
$R^{-4}\ln R$ and
$R^{-4}$, has not been evaluated previously.

The Bogoliubov equations for the excited states are rewritten in
hydrodynamic form and solved to incorporate the boundary layer to third
order in the boundary-layer thickness
$\delta\propto R^{-4/3}$.  This analysis provides a uniform extension of
the hydrodynamic normal modes found by Stringari\cite{stringari} beyond
the TF condensate throughout the classically forbidden region. The
lowest-order correction to the excitation frequencies has the form
$E_{nl}^\epsilon/R^4$ (namely, of order $\epsilon\equiv R^{-4}$).
Although a detailed calculation of the coefficient $E_{nl}^\epsilon$ appears
prohibitive, the shift in excitation frequencies due to finite-number effects
should be relevant for current experiments even when $\eta_0\gg 1$; both the
sum-rule approach~\cite{stringari} and numerical calculations~\cite{barry}
indicate that $E_{nl}^\epsilon$ increases dramatically with both $n$ and $l$.

\begin{acknowledgments}

The authors would like to thank S.~Stringari for stimulating discussions.
D.~L.~F. is indebted to Barry Schneider for valuable suggestions regarding
the numerical calculations.  This work has been partially supported by the
National Science Foundation under Grant No.~94-21888 (A.~L.~F.) and by the
Natural Sciences and Engineering Research Council of Canada and  the Ontario
Centre for Materials Research (D.~L.~F.).  In addition,  D.~L.~F.\ appreciates
the gracious hospitality of the Department of Physics at Stanford University.

\end{acknowledgments}

\appendix
\section{constants for the condensate}
\label{appendix1}

The constants $I$-$M$ appearing in Eqs. (\ref{normal})-(\ref{kinetic}) are
defined as follows:

\begin{eqnarray}
I&=&\int_{-\infty}^0dX(\Phi_0^2+X)+\int_0^{\infty}dX\,\Phi_0^2\\
J&=&\int_{-\infty}^{-1}dX\left(2\Phi_0^4+{5X\Phi_0^2\over 2}
+{X^2\over 2}-{3\over 8X}\right)
 +\int_{-1}^{\infty}dX\left(2\Phi_0^4+{5X\Phi_0^2\over 2}\right);
\label{J}\\
K&=&\int_{-\infty}^{-1}dX\left(2\Phi_0^4+{7X\Phi_0^2\over 2}
>+{3X^2\over 2}-{1\over 8X}\right)
+\int_{-1}^{\infty}dX\left(2\Phi_0^4+{7X\Phi_0^2\over 2}\right);\\
L&=&\int_{-\infty}^{-1}dX\left(\Phi_0^4-X^2-{1\over 2X}\right)
+\int_{-1}^{\infty}dX\,\Phi_0^4;\\
M&=&\int_{-\infty}^{-1}dX\left(\Phi_0^4+X\Phi_0^2-{1\over 4X}\right)
 +\int_{-1}^{\infty}dX\left(\Phi_0^4+X\Phi_0^2\right).\label{M}
\end{eqnarray}

\noindent Note that $K+L=J+M$. The integral $I$ can be evaluated analytically
by setting the lower limit to $X_0\rightarrow -\infty$ and integrating by
parts:
\begin{equation}
I=\lim_{X_0\to -\infty}{{X_0^2\over 2}-2\int_{X_0}^{\infty}dX\,X\Phi_0
\Phi_0'}.
\end{equation}

\noindent Multiplying Eq.\ (\ref{inner0}) by $\Phi_0'$ and integrating, one
readily obtains:
\begin{equation}
\int_{X_0}^{\infty}dX\,X\Phi_0\Phi_0'={1\over 2}\left[{\Phi_0'}^2-{\Phi_0^4
\over 2}\right]_{X_0}^{\infty}.
\end{equation}

\noindent The asymptotic behaviors (\ref{a_inner0}) and
$\Phi_0(X\rightarrow\infty)=0$ immediately give the result $I=0$.

The derivation of the expressions for $J$ and $K$ makes use of the relation

\begin{eqnarray}
\int_{-\infty}^{\infty}dX\,\Phi_0\Phi_1&=&-{23\over 16}-{3\over 4}
\int_{-\infty}^{-1}dX\,X^2\nonumber \\
& &\quad +\int_{-\infty}^{\infty}dX\left(2\Phi_0^4+{3\over 2}X\Phi_0^2\right)
\end{eqnarray}

\noindent which may be verified by integrating by parts, comparing with
Eqs.\ (\ref{inner0})-(\ref{a_inner1}), and employing the readily proved
identity
\begin{equation}
\int_{-\infty}^{\infty}dX\,{\Phi_0'}^2={1\over 2}-\int_{-\infty}^{\infty}dX
\left(\Phi_0^4+X\Phi_0^2\right).
\end{equation}

\noindent Furthermore, with the identity
\begin{equation}
2\int_{-\infty}^{\infty}dX\,X\Phi_0^2=-{1\over 6}-\int_{-\infty}^{-1}dX\,X^2
-\int_{-\infty}^{\infty}dX\,\Phi_0^4,
\end{equation}

\noindent which may be easily confirmed by integrating by parts and making use
of the governing equation (\ref{inner0}) for $\Phi_0(X)$, the expressions
(\ref{J})-(\ref{M}) are found to be related:

\begin{eqnarray}
J&=&{\case3/4}L-{\case5/{24}};\\
K&=&{\case1/4}L-{\case7/{24}};\\
M&=&{\case1/2}L-{\case1/{12}}.
\end{eqnarray}

\section{constants for the excitations}
\label{appendix2}

The constants appearing in the asymptotic expansions of the outer amplitudes
$F_{nl}^{0,\epsilon}$ and $G_{nl}^{0,\epsilon}$,
Eqs.~(\ref{F0_asympt})-(\ref{G1_asympt}), are explicitly written (where
$\rho_0(1)\equiv\rho_{nl}^0(1)$ and $E_0\equiv E_{nl}^0$):
\begin{eqnarray}
\rho_0'(1)&=& E_0^2\rho_0(1); \\
\rho_0''(1)&=& {\case1/2}\left[E_0^4-3E_0^2+l(l+1)\right]\rho_0(1); \\
\rho_0'''(1)&=& {\case1/{6}}\left[E_0^6-10E_0^4+5E_0^2\left(l(l+1)+5\right)
-13l(l+1)\right]\rho_0(1); \\
G_a^0&=& \rho_0(1); \\
G_b^0&=& \left[E_0^2-{\case1/2}\right]\rho_0(1); \\
G_c^0&=& {\case1/4}\left[E_0^4-5E_0^2+l(l+1)+1\right]\rho_0(1); \\
G_d^0&=& {\case1/{72}}\left[2E_0^6-29E_0^4+5E_0^2\left(2l(l+1)+19\right)
-35l(l+1)-9\right]\rho_0(1); \\
\nonumber \\
F_a^{\epsilon} &=& 0; \\
F_b^{\epsilon} &=& 0; \\
F_c^{\epsilon} &=& {\case1/{16}}\left[E_0^4-7E_0^2+3l(l+1)\right]\rho_0(1); \\
F_d^{\epsilon} &=& C_a^{\epsilon}; \\
F_e^{\epsilon} &=& C_b^{\epsilon}; \\
F_f^{\epsilon} &=& -{\case1/{16}}\left[E_0^4-E_0^2\left(4l(l+1)+3\right)
+5l(l+1)\right]\rho_0(1); \\
G_a^{\epsilon} &=& -{\case3/8}\rho_0(1); \\
G_b^{\epsilon} &=& -{\case1/8}\left[E_0^2-12\right]\rho_0(1); \\
G_c^{\epsilon} &=& -{\case1/{32}}\left[E_0^4-23E_0^2-11l(l+1)+66\right]\rho(1);
\\
G_d^{\epsilon} &=& -{\case1/{32}}\left[E_0^6-15E_0^4-E_0^2\left(3l(l+1)
-70\right)+8l(l+1)-66\right]\rho_0(1)+C_a^{\epsilon}; \\
G_e^{\epsilon} &=& C_b^{\epsilon}; \\
G_f^{\epsilon} &=& -{\case1/{16}}\left[E_0^4-E_0^2\left(4l(l+1)+3\right)
+5l(l+1)\right]\rho_0(1),
\end{eqnarray}

\noindent where $C_a^{\epsilon}$ and $C_b^{\epsilon}$ are constants of
integration.

\begin{figure}
\caption{The contribution of the kinetic energy in units of $\hbar\omega_0$ is
shown as a function of the universal scaling parameter $\eta_0=N_0a/d_0$. The
kinetic energy per particle is calculated numerically by direct integration of
the GP equation (solid bold) and analytically by the boundary-layer theory
(dashed bold), whose result is given in Eq.~(\ref{kinetic3}). The dashed thin
line is the difference between the exact total energy per particle (computed
numerically) and the full boundary-layer theory in Eq.~(\ref{Etot}). The solid
thin line corresponds to the similar difference keeping only the approximate
result of the TF theory [the first term of Eq.~(\ref{Etot})]. The total
energies per particle for a restricted range of $\eta_0$ are shown in the
inset for the exact (solid), boundary-layer (dashed), and TF (long dashed)
theories. On this restricted scale, the curves for the boundary-layer and
exact results coincide.}
\label{figure1}
\end{figure}

\begin{figure}
\caption{The (unnormalized) hydrodynamic amplitudes and condensate wave
function in the boundary layer are shown as a function of inner coordinate $X$.
Bold and thin lines correspond to results calculated numerically to order
$\delta$ and $\delta^2$, respectively. The universal parameter is
$\eta_0=1000$ giving $\delta\approx 0.06$, and therefore $X=-8$ corresponds
to $x\approx 0.5$. With units chosen appropriately, the velocity potential to
order $\delta$ (bold dot-dashed line) and the density perturbation [bold
solid line to order $\delta$, from Eq.~(\ref{rho})] coincide in the overlap
region $X\ll -1$; while the former diverges as $X\rightarrow\infty$, the
latter decays exponentially. The perturbations in the current density (dashed
lines) to order $\delta$ (bold) and $\delta^2$ (thin) are respectively linear
and quadratic in the overlap region and decay exponentially at large distances.
The results are presented for $(n,l)=(0,2)$; the overall sign of the inner
hydrodynamic amplitudes is odd in $n$, and the magnitude of the asymptotic
slope (for large negative $X$) increases with $l$. The inset shows the
various amplitudes in the bulk region as a function of outer coordinate $x$.
In the outer region, the velocity potential and density perturbation coincide
(shown as the solid line); the former matches smoothly with that from the
inner region to order $\delta^2$ (thin dot-dashed line). A perfect asymptotic
match of the outer current density $\propto x(1-x^2)$ (dashed line) to its
inner counterpart would require an inner expansion to order $\delta^3$.}
\label{figure2}
\end{figure}

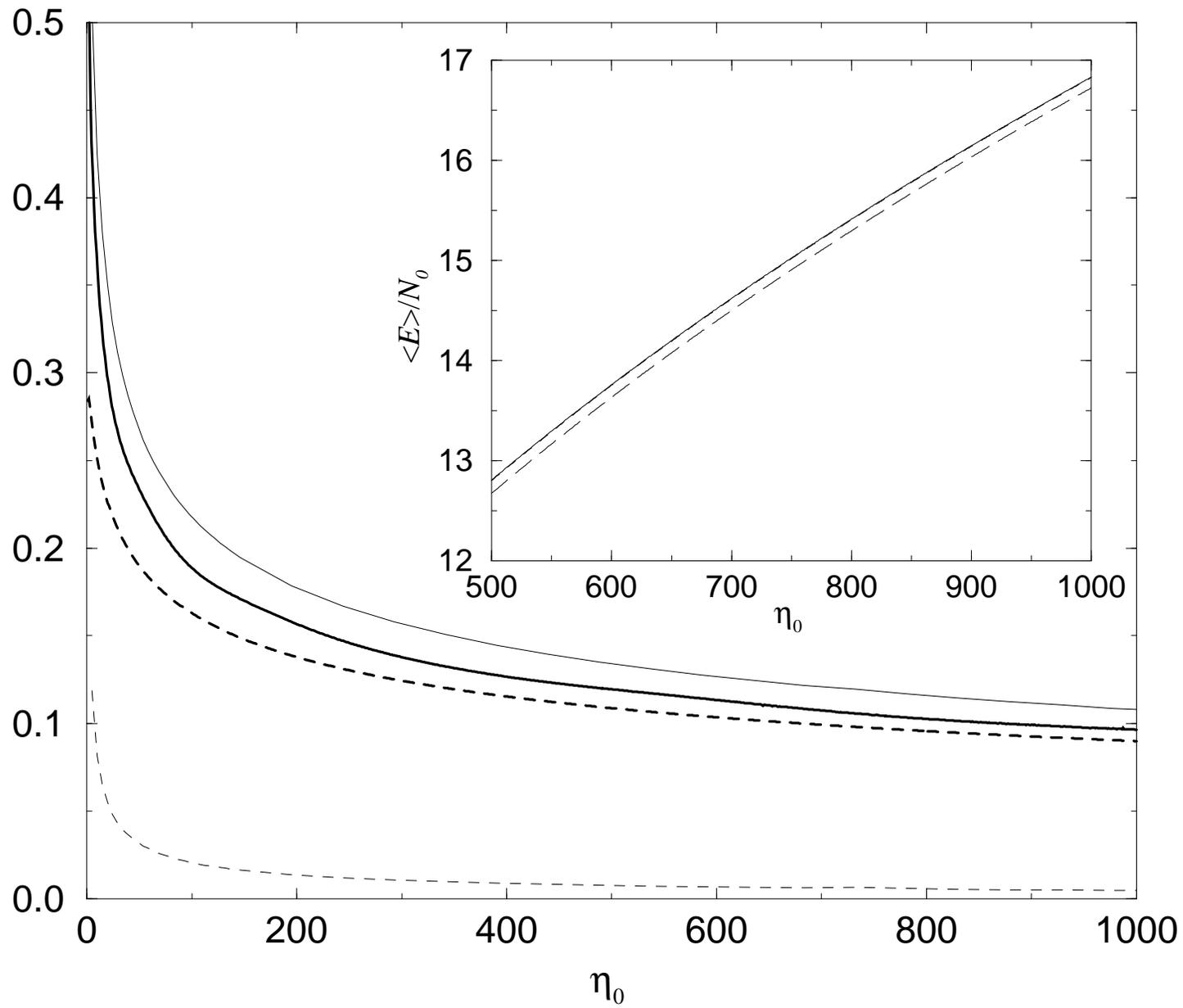
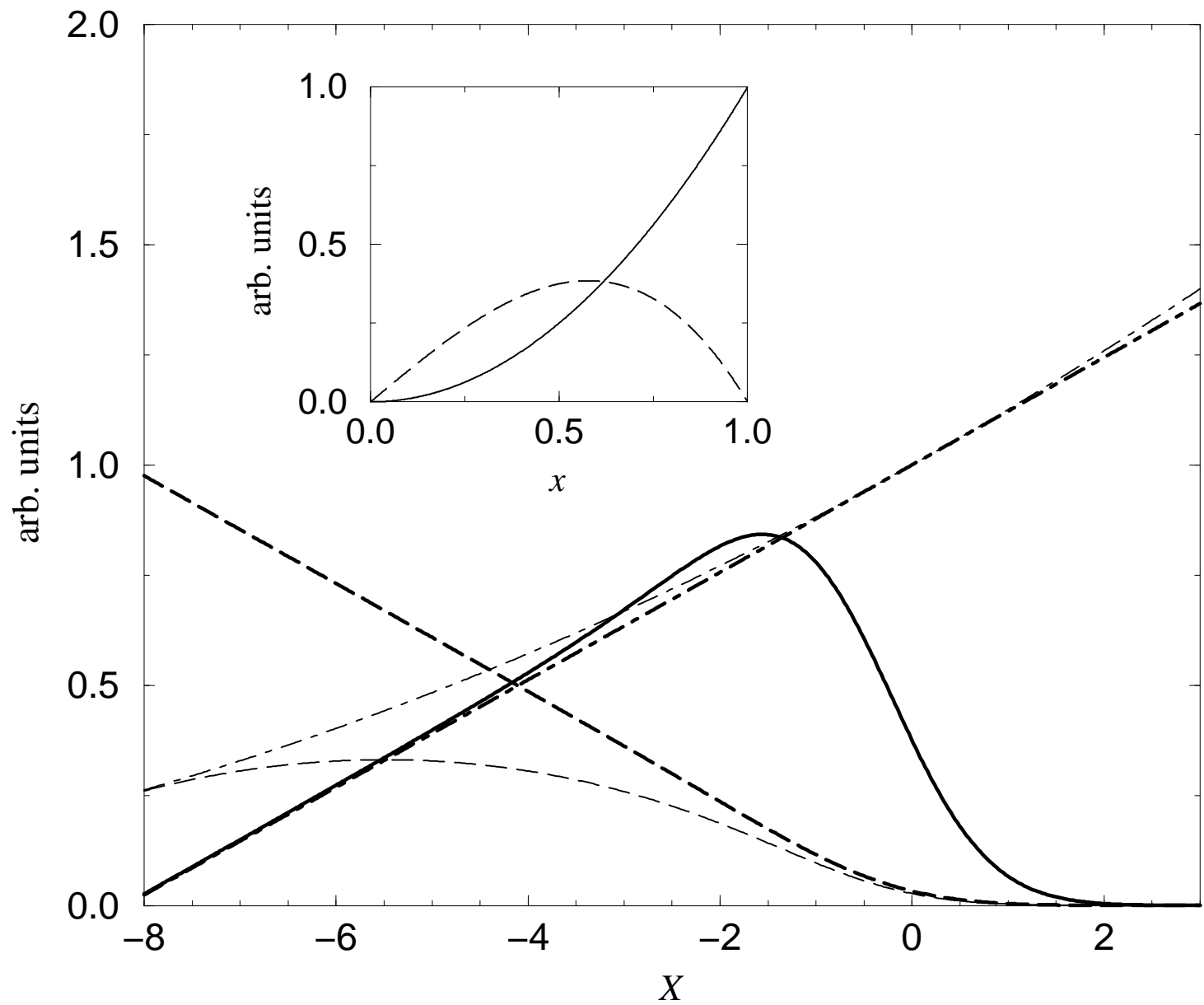
\end{document}